\documentclass{article}
\usepackage{amssymb}
\usepackage{amsmath}
\usepackage{amsthm}
\usepackage{amsfonts}
\usepackage{graphicx}

\newtheorem{theorem}{Theorem}

\newtheorem{definition}[theorem]{Definition}
\newtheorem{example}[theorem]{Example}

\newtheorem{lemma}[theorem]{Lemma}

\newtheorem{remark}[theorem]{Remark}

\renewenvironment{proof}[1][Proof]{\noindent\textbf{#1.} }{\
\rule{0.5em}{0.5em}}
\numberwithin{equation}{section}

\begin{document}

\title{On refined volatility smile expansion in the Heston model}

\author{Peter Friz\thanks{%
TU\ and WIAS\ Berlin, partially supported by MATHEON}, Stefan Gerhold\thanks{%
TU\ Wien, partially supported by the Austrian Federal Financing Agency and
the Christian-Doppler-Gesellschaft}, Archil Gulisashvili\thanks{%
Ohio University}, Stephan Sturm\thanks{%
TU\ Berlin, supported by MATHEON}}

\maketitle

\begin{abstract}
It is known that Heston's stochastic volatility model exhibits moment
explosion, and that the critical moment $s_+$ can be obtained by solving
(numerically) a simple equation. This yields a leading order expansion for
the implied volatility at large strikes: $\sigma _{BS}( k,T)^{2}T\sim \Psi
(s_+-1) \times k$ (Roger Lee's moment formula).
Motivated by recent ``tail-wing'' refinements
of this moment formula, we first derive a novel tail expansion for the
Heston density, sharpening previous work of Dr{\u{a}}gulescu and Yakovenko
[Quant.\ Finance 2, 6 (2002), 443--453], and  then show the validity of a refined expansion of the type $%
\sigma _{BS}( k,T) ^{2}T=( \beta _{1}k^{1/2}+\beta
_{2}+\dots)^{2}$, where all constants are explicitly known as functions
of~$s_+$, the Heston model parameters, spot vol and maturity~$T$. In the case of the
``zero-correlation'' Heston model such an expansion was derived by Gulisashvili and Stein
[Appl.\ Math.\ Optim. 61, 3 (2010), 287--315].
Our methods and results may prove useful beyond the Heston model: the entire
quantitative analysis is based on affine principles: at no point do we need
knowledge of the (explicit, but cumbersome) closed form expression of the
Fourier transform of $\log S_{T}$\ (equivalently: Mellin transform of~$S_{T}$%
); what matters is that these transforms satisfy ordinary differential equations
of Riccati type.
Secondly, our analysis reveals a new parameter (``\textit{critical slope}''%
), defined in a model free manner, which drives the second and higher order
terms in tail- and implied volatility expansions.
\end{abstract}

\section{Introduction}

The Heston model~\cite{He93} is one of the most popular stochastic volatility models used in
the financial industry. Furthering its understanding, and in particular the understanding of its implied volatility surface, is of particular interest in the light of the recent financial crisis: the volatility smile (underlying: SPX) did steepen after September 2008, then flattened again; it also steepened substantially after the flash crash in April of 2010 and has since flattened again\footnote{From a private communication with a derivative trader at a major investment bank.}. It is also worth recalling that the very existence of the volatility smile as we know it was triggered by the events of 1987.

This general motivation is complemented by an everyday question in the financial industry: how to (smoothly) extrapolate the smile seen in the market (typically a stepping stone towards the robust construction of a local volatility surface). Theorem 3 below contributes precisely in this direction and we derive new expansions for the implied volatility in the Heston model. Recall that its dynamics under the forward measure are given by 
\begin{align}
dS_{t}& =S_{t}\sqrt{V_{t}}dW_{t},\qquad S_{0}=1,  \notag \\
dV_{t}& =\left( a+bV_{t}\right) dt+c\sqrt{V_{t}}dZ_{t},\qquad V_{0}=v_{0}>0,
\label{E:H}
\end{align}%
where $a\geq 0$, $b\leq 0$, $c>0$, and $d\langle W,Z\rangle
_{t}=\rho dt$ with $\rho \in \lbrack -1,1]$. Observe that our choice $S_{0}=1
$, as well as zero drift, entails no loss of generality. As is well-known
(cf.\ \cite{AnPi07, BeFr08, FrKR10, KR08, LiMu07}), the Heston model, as many
other stochastic volatility models, exhibits \textit{moment explosion} in
the sense that%
\begin{equation*}
T^{\ast }( s) =\sup \left\{ t\geq 0:E[S_{t}^{s}]<\infty \right\} 
\end{equation*}%
is finite for~$s$ large enough.
(Here and throughout the paper, $E[\cdot]$ denotes the risk-neutral expectation.)
Differently put, for fixed maturity~$T$
there will be a (finite) \textit{critical moment}%
\begin{equation*}
s_+:=\sup \left\{ s\geq1 : E[S_{T}^{s}]<\infty \right\} .
\end{equation*}%
(In the Heston model, and many other affine stochastic volatility models,~$%
T^{\ast }$ is explicitly known. The critical moment, for fixed~$T$, is then
found numerically from $T^{\ast }( s_+) =T$.) A model free
result due to R.~Lee, known as moment formula (cf.~\cite{BeFrLe08, Le04a}; see
also~\cite{BeFr08, BeFr09, Fr10, Gu10}), then
yields%
\begin{equation}
\limsup_{k\rightarrow \infty }\sigma _{BS}(k,T)^{2}T=\Psi
( s_+-1) \times k  \label{LeeMomentFormula},
\end{equation}%
where $k=\log (K/S_{0})$ denotes the log-strike, $\sigma _{BS}$ the
Black-Scholes implied volatility, and%
\begin{equation*}
\Psi ( x) =2-4\bigl( \sqrt{x^{2}+x}-x\bigr) \in [ 0,2].
\end{equation*}%
We remark that, subject to some ``regularity'' of the moment blowup (fulfilled
in all practical cases; cf.~\cite{BeFr08}), the $\limsup $ can be replaced
by a genuine limit. Thus, the \textit{total implied variance} $\sigma
_{BS}( k,T) ^{2}T$ is asymptotically linear in~$k$ with slope~$\Psi
( s_+) $. (Similar results apply in the small strike limit $%
k\rightarrow -\infty$, but the focus of this paper is on $k\rightarrow \infty $.) 
 
Parametric forms of the implied volatility smile used in the industry
respect this behavior; a widely used parametrization is the following.

\begin{example}[Gatheral's SVI\ parametrization~\cite{Ga04}]
\label{ExSVI} For fixed~$T$, a parametric form of $\sigma _{BS}( k,T) ^{2}T$ is given by%
\begin{equation*}
k\mapsto \mathfrak{a}+\mathfrak{b}\left[ \left( -\mathfrak{m}+k\right) 
\mathfrak{r}+\sqrt{\left( -\mathfrak{m}+k\right) ^{2}+\mathfrak{s}}\right]
\equiv \mathrm{SVI}( k;\mathfrak{a},\mathfrak{b},\mathfrak{r},\mathfrak{m},%
\mathfrak{s}).
\end{equation*}%
An expansion for $k\rightarrow \infty $ yields%
\begin{align}
\mathrm{SVI}( k)  &= k\,\mathfrak{b}\left( 1+\mathfrak{r}\right) +\left( 
\mathfrak{a-bm}\left( 1+\mathfrak{r}\right) \right) +O( k^{-1}) ,
\notag \\
\sqrt{\mathrm{SVI}( k) } &= k^{\frac{1}{2}}\sqrt{\mathfrak{b}\left( 1+%
\mathfrak{r}\right) }+k^{-\frac{1}{2}}\frac{\left( \mathfrak{a-bm}\left( 1+%
\mathfrak{r}\right) \right) }{2\sqrt{\mathfrak{b}\left( 1+\mathfrak{r}%
\right) }}+O\bigl( k^{-\frac{3}{2}}\bigr),   \label{SqrtSVI}
\end{align}%
and we see that $\mathrm{SVI}(k) $ is asymptotically linear. Remark that
this parametrization is not ad-hoc but has been obtained by a $T\rightarrow
\infty $ analysis of the Heston smile; cf.~\cite{FoJaMi09} and~\cite{Ga04}.
\end{example}

Our main results are the following two theorems. Remark~\ref{rem:zero}
in Section~\ref{se:sp density} and formula~\eqref{E:a2zero}
in Section~\ref{se:smile} complement them by left-tail asymptotics.

\begin{theorem}
\label{T:main} For every fixed $T>0$, the distribution density~$D_{T}$ of
the stock price~$S_{T}$ in a correlated Heston model with $\rho \leq 0$
satisfies the following asymptotic formula: 
\begin{equation}
D_{T}(x) =A_{1}x^{-A_{3}}e^{A_{2}\sqrt{\log x}}\left( \log
x\right) ^{-3/4+a/c^{2}}\bigl(1+O((\log x)^{-1/2})\bigr)  \label{E:0}
\end{equation}%
as $x\rightarrow \infty $. The constants~$A_{3}$ and~$A_{2}$ are expressed
explicitly in terms of critical moment~$s_+$ and critical slope 
\begin{equation}
\sigma :=-\left. \frac{\partial T^{\ast }( s) }{\partial s}%
\right\vert _{s=s_+}  \label{DefSigma}
\end{equation}%
%and critical curvature
%\begin{equation*}
%\kappa := \left. \frac{\partial^2 T^{\ast }( s) }{\partial s^2}%
%\right\vert _{s=s_+}  
%\end{equation*}%
as
\begin{equation}\label{eq:A_i}
  A_{3} =s_++1 \qquad\text{and}\qquad A_{2} =2\frac{\sqrt{2v_{0}}}{c\sqrt{\sigma }}.
\end{equation}
An expression for~$A_1$ is presented in Remark~\ref{rem:A_1} below.
\end{theorem}
\begin{theorem}\label{thm:impl vol}
Under the assumptions of Theorem~\ref{T:main},
the Black-Scholes implied volatility admits the expansion
\begin{equation}
\sigma _{BS}( k,T) ^{2}T=\left( \beta _{1}k^{1/2}+\beta
_{2}+\beta _{3}\frac{\log k}{k^{1/2}}+O\left( \frac{1 }{%
k^{1/2}}\right) \right) ^{2}  \label{ImplVarExp}
\end{equation}%
as $k\to\infty$, where  
\begin{align*}
& \beta _{1}=\sqrt{2}\left( \sqrt{A_{3}-1}-\sqrt{A_{3}-2}\right),  \\
& \beta _{2}=\frac{A_{2}}{\sqrt{2}}\left( \frac{1}{\sqrt{A_{3}-2}}-\frac{1}{%
\sqrt{A_{3}-1}}\right),  \\
& \beta _{3}=\frac{1}{\sqrt{2}}\left( \frac{1}{4}-\frac{a}{c^{2}}\right)
\left( \frac{1}{\sqrt{A_{3}-1}}-\frac{1}{\sqrt{A_{3}-2}}\right) .
\end{align*}
\end{theorem}

\begin{remark} \rm
The restriction to $\rho \leq 0$ is (mathematically) not essential, but
allows to streamline the presentation. As is commonly noticed, this covers
essentially all practical applications of the Heston model. We also note
that, since $\left( a+bV_{t}\right) =-b\left( a/\left( -b\right)
-V_{t}\right) $, it can be helpful to think of $-b$ (resp. $\bar{v}=a/\left(
-b\right) $) as the speed of mean-reversion (resp.\ mean-reversion level) of the
Heston variance process.
\end{remark}

Let us draw attention to the main predecessors of this paper:
Dr{\u{a}}gulescu--Yakovenko~\cite{DrYa02} apply a saddle point argument to
deduce the leading order behavior of the density in the stationary variance
regime; essentially $D_{T}(x)
\approx x^{-A_{3}}$. Gulisashvili--Stein~\cite{GuSt09}
study the ``uncorrelated'' Heston model ($\rho =0$) and find the same
functional form as in~\eqref{E:0} and \eqref{ImplVarExp}, with (more involved)\ explicit expressions for $%
A\,_{i},\beta _{i}$. (Their method relies on representing call prices as average of Black-Scholes prices and does not apply when $\rho \ne 0$.) While it is easy to see that, in the case $\rho =0$, our expressions for $A_{3}$
agree, it is checked in Appendix~II (for the reader's peace of mind) that
our $A_{2}=2\frac{\sqrt{2v_{0}}}{c\sqrt{\sigma }}|_{\rho =0}$ coincides with
their expression for~$A_{2}$. In Appendix~III we present a numerical example that
shows the accuracy of our asymptotic formula for the density, and of the resulting
implied volatility expansion.

An interesting feature of our approach, somewhat in contrast to most
analytic treatments of the Heston model,\footnote{%
Exceptions include~\cite{Fa07, KR08}.} is that our entire quantitative
analysis is based on affine principles; at no point do we need knowledge of
the (explicit, but cumbersome) closed form expression of the Fourier
transform of $\log S_{T}$\, or, equivalently, the Mellin transform of~$S_{T}$.
(With one inconsequential exception, namely a simplification of the formula for the constant
factor~$A_1$.)
Instead, we are able to extract all the necessary information on the
transform by analyzing the corresponding Riccati equations near criticality,
using higher order Euler estimates.\footnote{%
See~\cite{FrVi08} for more information on the power of Euler estimates.} In
conjunction with a classical saddle point computation we then ``implement''
the Tauberian principle that the precise behavior  of the transformed
function near the singularity (the leading order of which is exactly
described by the critical slope!) contains all the asymptotic information
about the original function. At this heuristic level, we would expect that
the critical slope~$\sigma $, as defined in~\eqref{DefSigma}, is the key
quantity that drives the second and higher order terms in tail- and implied
volatility expansions of general stochastic volatility models (even in
presence of jumps). Back to a rigorous level, it appears that the key ingredients
of our analysis are applicable to general affine stochastic volatility
models (cf.~\cite{KR08}), and we will take up on this in future work.

The explicit constants $A_{i},\beta _{i}$ for $i=1,2,3$ in the above theorem
are clearly tied to the Heston model itself. In fact, it is the explicit
nature of how these constants depend on the Heston parameters $\left(
a,b,c,\rho \right) $, as well as spot vol~$v_{0}$ and maturity~$T$, that
furthers our understanding. Let us be explicit.\ It follows from
equation~\eqref{E:r1} below that $s_+=s_+( b,c,\rho
,T) $ does not depend on $a,v_{0}$ (equivalently: does not depend on $%
\bar{v},v_{0})$; furthermore $s_+( T) \rightarrow s_+( \infty ) \in ( 1,\infty ) $ as $T\rightarrow \infty 
$. Moreover, the critical slope is explicitly computable: $\sigma /T$ will
be seen to be an explicit fraction involving only $b,c,\rho $ and~$s_+$
but not $a,v_{0}$ (equivalently: $\bar{v},v_{0})$. We see furthermore that $%
1/\sigma =( T/\sigma )/T=O(1/T) $ as $T\rightarrow
\infty $. As a consequence of all this, we see that changes in spot vol~$%
\sqrt{v_{0}}$ are second order effects:~$\beta _{1}$ does not depend on~$%
\sqrt{v_{0}}$, whereas~$\beta _{2}$ depends linearly on it. Practically put,
we see that increasing spot vol allows to up-shift the smile (intuitively
obvious!) but does not affect its slopes at the extremes. We also note that changes in~$\bar{%
v}$ are not seen until looking at~$\beta _{3}$. No such information could be
extracted from~\eqref{LeeMomentFormula} and previous works.

Another application concerns the design of parametrizations of the implied
volatility: the SVI expansion~\eqref{SqrtSVI} is \emph{not} compatible
with the correct expansion~\eqref{ImplVarExp}; the latter has a constant
term, $\beta _{2}$, which is not present in~\eqref{SqrtSVI}. (We are
grateful to J.~Gatheral for pointing this out to us.) The solution to this
apparent contradiction (recall that SVI was obtained by a $T\rightarrow
\infty $ analysis of the Heston smile) is simply that $\beta _{2}\propto
A_{2}=O( \sigma ^{-1/2}) =O( T^{-1/2}) \to 0.$
In fact, this suggests that SVI type parametrizations could well benefit
from additional terms corresponding to such a $\beta _{2}$-term; essentially
accounting for the fact that $T\neq \infty $.

\section{Moment explosion in the Heston model}\label{se:moment}

\subsection{Heston model as an affine model and moment explosion}\label{se:He aff}

Consider the correlated Heston model given by~\eqref{E:H}, and set
$X_{t}=\log S_{t}$. From basic principles of affine diffusions (see, e.g.,~\cite%
{KR08}) we know that 
\begin{equation}
  \log E[ e^{sX_{t}}] =\phi(s,t)
    +v_{0}\psi(s,t), \label{eq:log E}
\end{equation}
where the functions~$\phi$ and~$\psi$ satisfy the following Riccati equations:
\begin{align}
\dot{\phi} &= F( s,\psi ),\,\,\phi(0) =0, \label{E:aa} \\
\dot{\psi} &= R( s,\psi ),\,\,\psi(0) =0,
\label{E:a}
\end{align}
with $F(s,v)=av$ and
$R(s,v)=\frac{1}{2}( s^{2}-s)+\frac{1}{2}c^2v^{2}+bv+s\rho cv$. In~\eqref{E:a}, 
$\dot{\phi}$ and~$\dot{\psi}$ are the partial derivatives with respect to~$t$ of the functions~$\phi$ and~$\psi$, respectively.
Our goal in Section~\ref{se:moment} is to identify the smallest singularity, $s=s_+$, of~\eqref{eq:log E},
and to analyze the asymptotic behavior of~\eqref{eq:log E} in its vicinity. The estimates found will be
put to use in Section~\ref{se:saddle}, where we perform the asymptotic inversion of the Mellin
transform~$E[e^{(u-1)X_t}]$ of the Heston model.
\begin{remark} \rm
The symbol~$s$ denotes a real parameter. The Riccati ODEs in~\eqref{E:aa} and~\eqref{E:a} are also valid when~$s$
is replaced by a complex parameter $u=s+iy$.
\end{remark}

Given $s\ge 1$, define the explosion time for the moment of order~$s$ by 
\[
T^{\ast }( s)=\sup \left\{
t\geq 0:E[ e^{sX_{t}}] <\infty \right\}.
\] 
An elementary computation gives%
\begin{equation*}
2c^2 \min_{\eta \in \lbrack 0,\infty ]}R\left( s,\eta \right) =-\left[ \left(
s\rho c+b\right) ^{2}-c^{2}\left( s^{2}-s\right) \right] =:-\Delta \left(
s\right).
\end{equation*}%
Let us also set $\chi ( s) =s\rho c+b$. A typical situation in
applications (a correlation parameter satisfying $\rho \leq 0$, and a non-zero mean reversion $%
b<0$) implies that~$\chi $ is negative for $s\ge 0$. We thus assume
in the sequel that 
\begin{equation*}
\chi(s) <0 \quad \text{for all} \quad s \geq 0.
\end{equation*}%
This assumption allows to use the following formula from~\cite[Theorem~4.2]{KR08}: 
\begin{equation}
T^{\ast }( s) =\left\{ 
\begin{array}{c}
+\infty \\ 
\int_{0}^{\infty }1/R( s,\eta ) d\eta%
\end{array}%
\right. \left. 
\begin{array}{c}
\text{if }\Delta ( s) \geq 0 \\ 
\text{if }\Delta ( s) <0%
\end{array}%
\right.
\label{E:r1}
\end{equation}

\begin{remark} \rm
\label{RmkExplosionTimeAndSlope}The integral in~\eqref{E:r1} can be represented as follows:
For $\Delta ( s) <0$, we have
\begin{equation}
T^{\ast }( s) =\frac{2}{\sqrt{-\text{ }\Delta( s) }}%
\left( \arctan \frac{\sqrt{-\text{ }\Delta ( s) }}{\chi (
s) }+\pi \right). \label{eq:T int}
\end{equation}%
The derivative 
\[
  \partial _{s}T^{\ast }=\int_{0}^{\infty }-\frac{%
  \partial _{s}R}{R^{2}}( s,\eta ) d\eta 
\] 
can be computed explicitly. Indeed, from~\eqref{eq:T int} we get
\begin{align}
\partial _{s}T^{*}(s)&=-T^{*}(s)\frac{2\rho c(s\rho c+b)-c^2(2s-1)}{2\Delta(s)} \nonumber \\
&\quad-\frac{\left[c^2(2s-1)-2\rho c(s\rho c+b)\right](s\rho c+b)+2\rho c\Delta(s)}
{\Delta(s)\left[(s\rho c+b)^2-\Delta(s)\right]}.
\label{DTstarExp}
\end{align}
\end{remark}

\subsection{Moment explosion}\label{se:moment expl}

For $t> 0$, let $s_{+}(t) \geq 1$ be the (generalized) inverse of the
(decreasing) function~$T^{\ast }(\cdot) $, that is%
\begin{equation*}
s_{+}( t) =\sup \left\{ s\geq 1:E[ e^{sX_{t}}] <\infty
\right\} .
\end{equation*}

\begin{definition}\label{def:s_+}
Given $T> 0$, we call%
\begin{equation*}
s_+:=s_{+}( T) =\sup \left\{ s\geq 1:E[ S_{T}^{s}%
] <\infty \right\}
\end{equation*}
the ``critical moment''. The quantities
\begin{equation*}
\sigma := -\partial _{s}T^{\ast }|_{s_+} \geq 0 \qquad \text{and} \qquad
\kappa:=\partial _{s}^2T^{\ast }|_{s_+}
\end{equation*}
are called the ``critical slope'' and the ``critical curvature'',
respectively. Note that~$s_+$, $\sigma$, and~$\kappa$ depend on~$T$.
\end{definition}
Since $T^{*}(s_+)=T$, formula~\eqref{DTstarExp} implies that
\begin{equation}
\sigma=-\frac{\partial T^{*}}{\partial s}(s_+)=\frac{R_1}{R_2},
\label{E:s}
\end{equation}
where
\begin{align*}
R_1&=Tc^2s_+\left(s_+-1\right)\left[c^2\left(2s_+-1\right)-2\rho c\left(s_+\rho c+b\right)\right] \\
&\quad-2\left(s_+\rho c+b\right)
\left[c^2\left(2s_+-1\right)-2\rho c\left(s_+\rho c+b\right)\right] \\
&\quad +4\rho c\left[c^2s_+\left(s_+-1\right)-\left(s_+\rho c+b\right)^2\right]
\end{align*}
and
\[
R_2=2c^2s_+\left(s_+-1\right)\left[c^2s_+\left(s_+-1\right)-\left(s_+\rho c+b\right)^2\right].
\]
\begin{remark} \rm
%We have already mentioned in Remark~\ref{RmkExplosionTimeAndSlope} that the functions $%
%T^{\ast }\left( \cdot \right) $ and $\partial _{s}T^{\ast }\left( \cdot \right) $ are
%available in closed form in the Heston model. 
The critical moment~$s_+$ can (and
in general: must) be obtained by a simple numerical root-finding procedure.
\end{remark}

Let $s\ge 1$. We know that $T^{\ast }( s) $ is the explosion time of~$\psi $.
On the other hand, using the Riccati ODE for~$\psi $, we see that%
\[
\left( 1/\psi \right) ^{\cdot }=-\frac{\dot{\psi}}{\psi ^{2}}
=-\frac{R( s,\psi ) }{\psi ^{2}}.
\]
Since $R(s,u)/u^2\to c^{2}/2$ as $u \rightarrow \infty$, we obtain
\begin{equation}
\psi (s,t) \sim \frac{1}{\frac{c^{2}}{2}\left( T^{\ast }(
s) -t\right) } \qquad \text{ as }t\uparrow T^{\ast}(s),
\label{E:t1}
\end{equation}%
uniformly on bounded subintervals of $[1,\infty)$. Next fix $T> 0$. Then we have $T=T^{\ast }( s_+) $
with $s_+=s_{+}( T) $. Since the function $T^{\ast }$ is continuously differentiable (and even $%
C^{2}$) in~$s$, we have 
\begin{align}
T^{\ast }( s) -T &= T^{\ast }( s) -T^{\ast }(s_+)  \notag \\
& = \left( s_+-s\right) \left( \sigma + O( s_+-s)
\right)  \label{TStarMinusTpropSStarMinusS} \\
&\sim \sigma \left( s_+-s\right) \qquad \text{ as }s\uparrow s_+,  \notag
\end{align}%
where $\sigma =-\partial _{s}T^{\ast }|_{s_+}$ is the critical slope. Hence

\begin{equation}
\psi \left( s,T\right) \sim \frac{2}{\left( s_+-s\right) c^{2}\sigma }%
\qquad \text{ as }s\uparrow s_+=s_{+}( T) .
\label{E:t2}
\end{equation}%
It follows from~\eqref{E:t1} and~\eqref{E:t2} that $\phi(s,t) =\int_{0}^{t}a\psi(s,\vartheta)d\vartheta $ 
has a logarithmic
blowup: % (which is also clear from the closed form of the mgf): 
\begin{equation*}
\phi(s,t) \sim -\frac{2a}{c^{2}}\log \left( T^{\ast }(
s) -t\right) \qquad \text{ as }t\uparrow T^{\ast }( s) ;
\end{equation*}%
or%
\begin{equation*}
\phi (s,T) \sim -\frac{2a}{c^{2}}\log \left( \left( s^{\ast
}-s\right) \sigma \right) \qquad \text{ as }s\uparrow s_+=s_{+}(T).
\end{equation*}
The following lemma refines these asymptotic results.
\begin{lemma}\label{le:dev}
For every $T> 0$ and for $s\uparrow s_+=s_{+}(T)$, the following formulas hold: 
\begin{align}
\psi(s,T) &= \frac{2}{\left( s_+-s\right) c^{2}\sigma }-%
\frac{b+s_+\rho c}{c^{2}}- \frac{\kappa}{c^2\sigma^2}+O(s_+-s),  \label{PsiTBehaviour} \\
\phi (s, T) &= \frac{2a}{c^2}\log\frac{1}{s_+-s} + \frac{2a}{c^2} \log\frac{T}{\sigma} \notag \\
    &\qquad + a \int_0^T \left( \psi(s_+,\vartheta) - \frac{2}{c^2(T-\vartheta)} \right) d\vartheta
    + O(s_+-s). \label{PhiTBehaviour}
\end{align}
\end{lemma}
\begin{proof}
The idea is to use (second order)\ Euler estimates for the Riccati ODEs near
criticality; this yields the limiting behavior of $\psi (s, t)$ and $\phi
(s, t) $ as $t\uparrow T^{\ast }( s)$, and we complete the proof 
using~\eqref{TStarMinusTpropSStarMinusS}. More precisely, let us introduce
time-to-criticality $\tau =T^{\ast }( s) -t$, and set $\hat{\psi}%
(s,\tau ) =\psi (s,T^{\ast }( s) -\tau ) $.
Observe that $1/\hat{\psi}(s, 0) =0$ and%
\begin{align*}
( 1/\hat{\psi}) ^{\cdot }&=-\frac{( \hat{%
\psi}) ^{\cdot }}{\hat{\psi}^{2}}=\frac{1}{\hat{\psi}^{2}}R( s,\hat{\psi})  \\
&=\frac{c^{2}}{2}+\frac{b+s\rho c}{\hat{\psi}}+\frac{s^{2}-s}{2\hat{\psi}%
^{2}}=W(s,1/\hat{\psi}),
\end{align*}%
where $W(s,u)=\frac{c^2}{2}+(b+s\rho c)u+\frac{s^2-s}{2}u^2$. 
A higher order Euler scheme for this ODE yields%
\begin{equation*}
( 1/\hat{\psi}) (s, \tau ) =( 1/\hat{\psi})
(s, 0) +W(s, 0) \tau +W(s, 0) W^{\prime }(s, 0)\tau
^{2}/2+o( \tau ^{2})
\end{equation*}%
as $\tau\rightarrow 0$ and~$s$ stays in a bounded interval. 
Since $W(s, 0) =\frac{c^{2}}{2}$ and $W^{\prime }(s, 0)=b+s\rho c$, we obtain
\begin{align*}
1/\hat{\psi}(s, \tau )  &= \frac{c^{2}}{2}\tau \left( 1+\frac{%
b+s\rho c}{2}\tau +O( \tau ^{2}) \right)  \\
&= \frac{c^{2}}{2}\tau \left( 1-\frac{b+s\rho c}{2}\tau +O( \tau
^{2}) \right) ^{-1}.
\end{align*}%
It follows that
\begin{align}
\hat{\psi}\left(s, \tau \right)  &= \frac{1}{\frac{c^{2}}{2}\tau }\left( 1-%
\frac{b+s\rho c}{2}\tau +O( \tau ^{2}) \right) \notag \\
&= \frac{2}{c^{2}\tau }-\frac{b+s\rho c}{c^{2}}+O( \tau ) \label{eq:psi hat}
\end{align}%
as $\tau =T^{\ast }( s) -t\downarrow 0$.
Note that
\begin{align*}
  \frac{1}{\tau} &= \left(\sigma(s_+-s) + \tfrac12 \kappa(s_+-s)^2 + O((s_+-s)^3)\right)^{-1}  \\
  &= \frac{1}{\sigma(s_+-s)} - \frac{\kappa}{2\sigma^2} + O(s_+-s). 
\end{align*}
Hence we obtain
\begin{equation*}
\psi \left(s, T\right) =\frac{2}{c^{2}\sigma \left( s_+-s\right)  }-\frac{%
b+s_+\rho c}{c^{2}} - \frac{\kappa}{c^2\sigma^2} + O( s_+-s)
\end{equation*}%
as $s\uparrow s_+=s_{+}( T) .$ 
%Finally, we can integrate the expansion%
%\begin{equation*}
%\hat{\psi}\left(s, \tau \right) =\frac{2}{c^{2}\tau }+O( 1) \qquad \text{
%as }\tau \downarrow 0
%\end{equation*}%
%termwise, which gives
%\begin{equation*}
%\phi \left(s, T^{\ast }( s) -\tau \right) =-\frac{2a}{c^{2}}\log
%\tau +O( \tau ) \qquad \text{ as }\tau \downarrow 0.
%\end{equation*}%
For the expansion of $\phi(s,t) =\int_{0}^{t}a\psi(s,\vartheta)d\vartheta $, we find
\begin{align}
  \phi(s,t) &= a \int_0^t \left( \psi(s,\vartheta) - \frac{2}{c^2(T^*(s)-\vartheta)} \right) d\vartheta
    + \frac{2a}{c^2} \int_0^t \frac{1}{T^*(s)-\vartheta} d\vartheta \notag \\
  &= \frac{2a}{c^2} \log \frac{1}{T^*(s)-t} + \frac{2a}{c^2} \log T^*(s) + a \int_0^t \left( \psi(s,\vartheta) -   
    \frac{2}{c^2(T^*(s)-\vartheta)} \right) d\vartheta \notag \\
  &= \frac{2a}{c^2} \log \frac{1}{T^*(s)-t} + \frac{2a}{c^2} \log T^*(s) \notag \\
  & \qquad + a \int_0^{T^*(s)} \left( \psi(s,\vartheta) -   
    \frac{2}{c^2(T^*(s)-\vartheta)} \right) d\vartheta  + O(T^*(s)-t). \label{eq:phi proof}
\end{align}
To see the last equality, note that the integrand of
\[
  \int_t^{T^*(s)} \left( \psi(s,\vartheta) -   
    \frac{2}{c^2(T^*(s)-\vartheta)} \right) d\vartheta  = O(T^*(s)-t)
\]
has an expansion resulting from~\eqref{eq:psi hat}, which may be integrated termwise~\cite{deBr81}.
It now suffices to use~\eqref{TStarMinusTpropSStarMinusS} and~\eqref{eq:phi proof} to see that, as $%
s\uparrow s_+=s_{+}( T) ,$ formula~\eqref{PhiTBehaviour} holds.
\end{proof}
\begin{remark}\label{R:vazh} \rm It follows easily from the proof that
Lemma~\ref{le:dev} also holds as~$s$ tends to~$s_+$ in the complex plane,
provided that~$\Re(s)<s_+$.
%More precisely,
%$$
%\psi \left(u,T\right)=\frac{2}{\left( s_+-u\right) c^{2}\sigma }+
%\frac{b+u\rho c}{c^{2}}+O(\left|s_+-u\right|)
%$$
%and
%$$
%\phi (u,T)=-\frac{2a}{c^{2}}\log (s_+-u)-\frac{2a}{c^{2}}\log \sigma
%+O(\left|s_+-u\right|).
%$$
\end{remark}

%Stefan: removed the following remark, as it is unclear,
%  at least here, before we have talked about saddles.
%
%\begin{remark} \rm
%The blowup of $\phi \left(s, T\right) +v_{0}\psi \left(s, T\right) $ as $%
%s\rightarrow s_+=s_{+}\left( T\right) $ is obviously dominated by the $%
%1/\left( s_+-s\right) $ blowup of~$\psi $. When computing the
%(approximate)\ saddle point, the contribution of~$\phi $ can be ignored. However, it
%does enter the local expansion around the saddle point.
%\end{remark}

\section{Mellin inversion via saddle point method}\label{se:saddle}

Our proof of Theorem~\ref{T:main} proceeds by an asymptotic analysis
of $E[ e^{\left( u-1\right) X_{T}}]$,
where~$u$ is complex. This is the Mellin transform of the density of~$S_{T}$.
As noted in Section~\ref{se:He aff} above, we can represent it in terms of the
functions~$\phi$ and~$\psi$ appearing in the Riccati ODEs: 
\[
\log E[e^{\left( u-1\right) X_{T}}] =\phi \left(u-1,T\right)+v_{0}\psi
\left(u-1,T\right).
\] 
The density can be recovered using the Mellin inversion formula, that is
\begin{equation}
D_T(x)=\frac{1}{2\pi i}\int_{s-i\infty }^{s+i\infty
}e^{-uL+\phi \left(u-1, T\right) +v_{0}\psi \left(u-1,T\right) }du,
\label{E:i1}
\end{equation}
where $L=\log x$, provided that~$s$ is in the fundamental strip, $s\in \left( s_{-}(T)
,s_{+}(T) \right) $.

\begin{remark} \rm
The integral in~\eqref{E:i1} exists, since its integrand decays exponentially at~$\pm i\infty$ (see Lemma~\ref%
{le:exptailWithODE} in Appendix~I). Moreover,
if $u-1$ is imaginary, then the characteristic function of the random variable 
$X_{T}=\log \left(S_{T}\right) $ decays exponentially.
It follows that~$X_{T}$ (and therefore~$S_{T}$) admits a smooth density.
Since~$S_{T}$ is (a component) of a locally elliptic diffusion with smooth
coefficients, this can also be seen employing classical stochastic or PDE\
methods (see~\cite{deMa09} for some recent advances in this direction).
\end{remark}

We will deduce the asymptotics of~\eqref{E:i1} by the
saddle point (or steepest descent) method~\cite{deBr81,FlSe09}.
The main idea is to deform the contour of integration into a
path of steepest descent from a saddle point of the integrand. In cases where the method can
be applied successfully, the saddle becomes steeper and more pronounced as the
parameter ($x$ in our case)
increases. We then replace the integrand with a local expansion around the
saddle point. The resulting integral, taken over a small part of the contour
containing the saddle point, is easy to evaluate asymptotically. Finally, it suffices to
show that the tails of the original integral are
negligible, in order to establish the asymptotics of the original integral.
Our treatment bears similarities to Taylor expansions studied by Wright~\cite{Wr32} and to the saddle point analysis of certain Lindel\"of
integrals~\cite{FlGeSa10}. The type of the pertinent singularity (exponential of a pole)
is the same in all cases.

\subsection{Finding the saddle point}

A (real) saddle point of the integrand in formula~\eqref{E:i1} can be found by equating its
derivative to zero. Since it usually suffices to calculate an approximate saddle point, we note
that Lemma~\ref{le:dev} and Remark~\ref{R:vazh} imply the following expansion, as
$u\to u^*:=s_++1=A_3$ with $\Re(u)<u^*$:
\begin{equation}\label{eq:u* expan}
  \phi(u-1,T) + v_0 \psi(u-1,T)=\frac{\beta^2}{u^*-u} + \frac{2a}{c^2} \log \frac{1}{u^*-u}
  + \Gamma + O(u^*-u),
\end{equation}
where we put $\beta^2=2v_0/c^2\sigma$ and
\begin{equation}\label{eq:def Gamma}
  \Gamma = -v_{0}\left(\frac{b+s_+\rho c}{c^{2}} + \frac{\kappa}{c^2\sigma^2} \right) + \frac{2a}{c^{2}}
    \log\frac{T}{\sigma} + a \int_0^T \left( \psi(s_+,\vartheta) - \frac{2}{c^2(T-\vartheta)} \right) d\vartheta.
\end{equation}
Retaining only the dominant term of~\eqref{eq:u* expan}, we get the approximate saddle point equation:
\[
  \left[x^{-u} \exp\left( \frac{\beta^2}{u^*-u} \right)\right]^{\prime} = 0,
\]
or equivalently,
\[
  -L + \frac{\beta^2}{(u^*-u)^2} = 0.
\]
The solution to the previous equation,
\[
  \hat{u} = \hat{u}(x) := u^* - \beta L^{-1/2},
\]
is the approximate saddle point of the integrand.

\subsection{Local expansion around the saddle point}

Our next goal is to expand the function $\phi (u-1,T)+v_{0}\psi (u-1,T)$ at the point~$u=\hat{u}$. Put $u=\hat{u}%
+iy$, and recall that we use the following notation: $\sigma= -\partial _{s}T^{\ast }|_{s_+}$ and $L=\log x$.
Since the (approximate) saddle point~$\hat{u}$ approaches~$u^*$ as $L\to\infty$, we may find
the expansion of the integrand using~\eqref{eq:u* expan}. To make the expansion valid uniformly w.r.t.\ the new
integration parameter~$y$, we confine~$y$ to the following small
interval: 
\begin{equation}
|y|<L^{-\alpha },\qquad \tfrac{2}{3}<\alpha<\tfrac{3}{4}.  \label{eq:y range}
\end{equation}%
The choice of the upper bound on~$\alpha $ in~\eqref{eq:y range} will be clear from
the tail estimates obtained in Appendix~I.
Since $u^*-u = \beta L^{-1/2} - iy$, we have
\begin{align}
  \frac{1}{u^*-u} &= \beta^{-1}L^{1/2} (1 - i\beta^{-1}L^{1/2}y)^{-1}\nonumber  \\
  &= \beta^{-1} L^{1/2} (1 + i \beta^{-1}L^{1/2}y - \beta^{-2}Ly^2 + O(L^{3/2-3\alpha})) \nonumber  \\
  &= \beta^{-1} L^{1/2} + i \beta^{-2}Ly - \beta^{-3}L^{3/2}y^2 + O(L^{2-3\alpha}). 
\label{eq:recipr expan}
\end{align}
It follows that
\begin{align*}
  \log \frac{1}{u^*-u} &= \log\left[\beta^{-1}L^{1/2} (1 + O(L^{1/2-\alpha}) )\right] \\
  &= \frac12 \log L - \log \beta + O(L^{1/2-\alpha}).
\end{align*}
Next, plugging the previous expansions, with $u=\hat{u}+iy$, into~\eqref{eq:u* expan}, we obtain 
the following asymptotic formula:
\begin{align}
 &\phi(\hat{u}-1+iy,T)+ v_0 \psi(\hat{u}-1+it,T) \nonumber \\
&=\beta L^{1/2} + \frac{a}{c^2} \log L + i L y - \beta^{-1}L^{3/2}y^2
  - \frac{2a}{c^2}\log\beta + \Gamma + O(L^{2-3\alpha}).
\label{E:touse}
\end{align}
%For $u=\hat{u}+iy$, we have
%$x^{-u} = x^{-\hat{u}} \exp(- i Ly)$.
%Therefore, the following :
%\begin{multline}\label{eq:sp expan}
%  x^{-\hat{u}-iy} \exp(\phi(\hat{u}-1+iy,T) + v_0 \psi(\hat{u}-1+iy),T)   \\
%    = x^{-\hat{u}} \exp\left(\beta L^{1/2} + \frac{a}{c^2} \log L  - \beta^{-1}L^{3/2}y^2
%  - \frac{2a}{c^2}\log\beta + \Gamma\right) (1 + O(L^{2-3\alpha})).
%\end{multline}

\subsection{Saddle point approximation of the density}\label{se:sp density}

For the sake of simplicity, we will first obtain formula~\eqref{E:0} with a weaker error estimate 
$O((\log x)^{-1/4+\varepsilon})$, where $\varepsilon> 0$ is arbitrary. Then it will be explained
how to get the stronger estimate $O((\log x)^{-1/2})$.

We shift the contour in the Mellin inversion formula~\eqref{E:i1}
through the saddle point~$\hat{u}$, so that%
\begin{align}
D_T(x)&=\frac{1}{2\pi i}\int_{\hat{u}-i\infty }^{%
\hat{u}+i\infty }e^{-uL+\phi \left(u-1,T\right) +v_{0}\psi \left(
u-1,T\right) }du  \label{eq:integral1} \\
&=x^{-\hat{u}}\frac{1}{2\pi}\int_{-\infty }^{\infty }e^{-iyL+\phi
\left(\hat{u}+iy-1,T\right) +v_{0}\psi \left( \hat{u}+iy-1,T\right) }dy.
\label{eq:integral}
\end{align}%
The term
\[
  x^{-\hat{u}} \approx x^{-u^*} = x^{-A_3}
\]
will yield the leading-order decay in~\eqref{E:0};
its exponent corresponds to the \emph{location} of the dominating singularity of the Mellin transform.
The lower order factors are dictated by the \emph{type} of the singularity at $u=u^*$, to be unveiled in what follows.

The ``tail'' of the last integral in~\eqref{eq:integral}, corresponding to $|y|>L^{-\alpha}$,
%, to which the local expansion~\eqref{eq:sp expan}
%can not be applied,
can be estimated using Lemma~\ref{le:tail} (see Appendix I). Therefore,
\begin{align*}
D_T(x)&=x^{-\hat{u}}\frac{1}{2\pi}\int_{-L^{-\alpha }}^{L^{-\alpha
}}e^{-iyL+\phi \left(\hat{u}+iy-1,T\right) +v_{0}\psi \left(\hat{u}
+iy-1,T\right)}dy \\
&\quad+x^{-A_{3}}\exp \left( 2\beta L^{1/2}-\beta
^{-1}L^{3/2-2\alpha }+O(\log L)\right).
\end{align*}%
Next, using~\eqref{E:touse} and the equality $x^{-\hat{u}}\exp(\beta L^{1/2})=x^{-u^{*}}
\exp(2\beta L^{1/2})$, we obtain
%the local expansion~\eqref{eq:sp expan}. 
\begin{align}
&D_T(x)=\frac{\exp \left( \Gamma \right) }{2\pi }
x^{- u^{\ast } }e^{2\beta L^{1/2}}
\beta^{-2a/c^2} L^{a/c^2}
\int_{-L^{-\alpha }}^{L^{-\alpha }}\exp \left(
-\beta^{-1}L^{3/2}y^{2}\right) dy \nonumber \\
&\quad\times (1+O(L^{2-3\alpha }))
+x^{-A_{3}}\exp \left( 2\beta L^{1/2}-\beta^{-1}L^{3/2-2\alpha }+O(\log L)\right).  \label{eq:expansion inserted}
\end{align}
Evaluating the Gaussian integral, we get
\begin{align}
&\int_{-L^{-\alpha }}^{L^{-\alpha }}\exp (-\beta ^{-1}L^{3/2}y^{2})dy =\beta
^{1/2}L^{-3/4}\int_{-\beta ^{-1/2}L^{3/4-\alpha }}^{\beta
^{-1/2}L^{3/4-\alpha }}\exp (-w^{2})dw \nonumber \\
& \sim \beta ^{1/2}L^{-3/4}\int_{-\infty }^{\infty }\exp (-w^{2})dw=\sqrt{%
\pi }\beta ^{1/2}L^{-3/4}.
\label{E:eq2}
\end{align}%
Here we use the fact that the tails of the Gaussian integral are exponentially small in~$L$. Taking into account~\eqref{eq:expansion inserted} 
and~\eqref{E:eq2}, we can compare 
the main part of the asymptotic expansion and the two error terms: 
\begin{align*}
&\mathrm{const}\times x^{-A_{3}}L^{a/c^{2}-3/4}\exp (2\beta L^{1/2}) & & \text{(main part)} \\
&x^{-A_{3}}L^{a/c^{2}-3/4}\exp(2\beta L^{1/2})\ O(L^{2-3\alpha }) & & \text{%
(error from local expansion)} \\
&x^{-A_{3}}\exp (2\beta L^{1/2}-\beta ^{-1}L^{3/2-2\alpha
}+O(\log L)) & & \text{(error from tail estimate)}
\end{align*}%
Since $2-3\alpha< 0$, the expression on the second line is asymptotically smaller than the main part. 
In addition, since $3/2-2\alpha >0$, the quantity $\exp (-%
\beta ^{-1}L^{3/2-2\alpha })$ decays faster than any power of~$L$. This shows that the expression on the 
third line is negligible in comparison with the error term in the local expansion. 
Hence, it suffices to keep only the error term resulting
from the local expansion. As a result, the error term in the asymptotic formula for~$D_T$ is
$O(L^{2-3\alpha })=O(L^{-1/4+\varepsilon })$. (Take~$\alpha $ close to~$%
\tfrac{3}{4}$.) More precisely, using~\eqref{eq:expansion inserted} and~\eqref{E:eq2}, 
we get the following formula:
\begin{align}
&D_T(x)
%=x^{-\left( s_++1\right) }e^{2\beta L^{1/2}}\left( \frac{1}{X\sigma }
%\right) ^{\frac{2a}{c^{2}}}\sqrt{\pi }\beta ^{-1}X^{3/2}
%(1+O(L^{2-3\alpha })) \nonumber \\
%&=\left[ \left( \frac{1}{\sigma }\right) ^{\frac{2a}{c^{2}}}\sqrt{\pi }
%\beta ^{-1}\right] \text{ }x^{-\left( s_++1\right) }e^{2\beta
%L^{1/2}}X^{3/2-2a/c^{2}}(1+O(L^{2-3\alpha })) \nonumber \\
=\left[\frac{\exp \left( \Gamma \right)}{2\pi }\sqrt{\pi }
\beta ^{1/2-2a/c^{2}}\right] x^{-\left( s_++1\right) }e^{2\beta
L^{1/2}}L^{-3/4+a/c^{2}} \nonumber \\
&\quad\times(1+O(L^{-1/4+\varepsilon })).
\label{E:fin}
\end{align}

It follows from~\eqref{E:fin} that formula~\eqref{E:0}, with a weaker error estimate,
holds for the correlated Heston model of our interest.

\begin{remark}\label{rem:A_1} \rm
  The integral on the
  right-hand side of~\eqref{eq:def Gamma} can be easily calculated from the closed form
  expression~\cite{RoFeUt09,He93} of~$\psi$.
  %Note by Stefan: I actually did not calculate the integral, but used the explicit expression of phi
  %  to get the constant term of phi's expansion.
  By~\eqref{E:fin} , we thus obtain the explicit expression
  \begin{align*}
    A_1 &= \frac{1}{2\sqrt{\pi }} \left( 2v_{0}\right) ^{1/4-a/c^{2}}c^{2a/c^{2}-1/2}\sigma^{-a/c^{2}-1/4} \notag \\
    & \qquad \times  \exp\left( -v_0 \left( \frac{b+s_+ \rho c}{c^2}+\frac{\kappa}{c^2\sigma^2} \right) 
     -\frac{aT}{c^2}(b+c\rho s_+) \right) \notag \\
    & \qquad \times
    \left(\frac{2\sqrt{b^2+2bc\rho s_++c^2 s_+(1-(1-\rho^2)s_+)}}{c^2 s_+(s_+-1)
    \sinh \frac12\sqrt{b^2+2bc\rho s_++c^2 s_+(1-(1-\rho^2)s_+)} } \right)^{2a/c^2}
  \end{align*}
  for the constant factor in~\eqref{E:0}.
\end{remark}

Our next goal is to show how to obtain the relative error $O((\log x)^{-1/2})$
in formula~\eqref{E:0}. Taking two more
terms in the expansion~\eqref{eq:recipr expan} of~$1/(u^*-u)$,
we get
\begin{align*}
  &\frac{1}{u^*-u}= \beta^{-1}L^{1/2} (1 - i\beta^{-1}L^{1/2}y)^{-1} \\
  &= \beta^{-1} L^{1/2} (1 + i \beta^{-1}L^{1/2}y - \beta^{-2}Ly^2 - i \beta^{-3}L^{3/2}y^3
  + \beta^{-4}L^2y^4 +  O(L^{5/2-5\alpha})) \\
  &= \beta^{-1} L^{1/2} + i \beta^{-2}Ly - \beta^{-3}L^{3/2}y^2
  - i \beta^{-4} L^2y^3 + \beta^{-5}L^{5/2}y^4 + O(L^{3-5\alpha}).
\end{align*}
%The~$y^5$-term yields an error of $O(L^{3-5\alpha})=O(L^{-3/4+\varepsilon})$. 
Expanding the logarithm, we obtain
\begin{align*}
  \log \frac{1}{u^*-u} &=\log(\beta^{-1}L^{1/2} (1 + i \beta^{-1}L^{1/2}y - \beta^{-2}Ly^2+ O(L^{3/2-3\alpha}) ) ) \\
  &= \frac12 \log L - \log \beta + i \beta^{-1}L^{1/2}y -\tfrac12 \beta^{-2}Ly^2 + O(L^{3/2-3\alpha}).
\end{align*}
%Next, plugging the previous two expansions with $u=\hat{u}+iy$ into~\eqref{eq:u* expan},  
%we get the following improvement of formula~(\ref{E:touse}): 
%\begin{align*}
%&\phi(\hat{u}-1+iy,T)+ v_0 \psi(\hat{u}-1+it,T)  \\
%&=\beta L^{1/2} + \frac{a}{c^2} \log L+ i \beta^{-1}\frac{2a}{c^2}L^{1/2}y + i L y 
%- \beta^{-1}L^{3/2}y^2-i\beta^{-2} L^2y^3  \\
%&\quad+ \beta^{-3}L^{5/2}y^4
% - \frac{2a}{c^2}\log\beta + \Gamma + O(L^{1-2\alpha}).
%\end{align*}
We insert these two expansions into~\eqref{eq:u* expan}
to obtain a refined expansion of the integrand:
\begin{align}
 & x^{-\hat{u}-iy}\exp\left(\phi(\hat{u}-1+iy,T)+ v_0 \psi(\hat{u}-1+it,T)\right) \nonumber \\
&=x^{-u^*} \exp\left(2\beta L^{1/2} + \frac{a}{c^2} \log L - \beta^{-1}L^{3/2}y^2
  - \frac{2a}{c^2}\log\beta + \Gamma\right) \nonumber \\
 &\quad\left( 1 + c_1 L^2 y^3 + c_2 L^{5/2} y^4
  +c_3 L^{1/2}y + c_4 Ly^2 + c_5 L^{-1/2} + O(L^{-3/4+\varepsilon}) \right),
\label{E:refi}
\end{align}
for some constants~$c_1,\dots,c_5$. Note that the terms with~$c_1$ and~$c_2$ come
from $(u^*-u)^{-1}$, those involving~$c_3$ and~$c_4$ from $\log(u^*-u)^{-1}$,
and the one with~$c_5$ from~$u^*-u$. 
(To be precise, we have used that the $O()$-term
in~\eqref{eq:u* expan} is of the form $c(u^*-u) + O((u^*-u)^2)$, as is easily seen by a third
order Taylor expansion along the lines of Section~\ref{se:moment expl}.)

We will next reason as in the proof of the weaker error estimate. The main term and the error term from the 
tail estimate remain the same. The error term from the local expansion can be obtained as follows:
Integrate the functions on both sides of formula~\eqref{E:refi} and take into account that
\[
\int_{L^{-\alpha}}^{L^{-\alpha}}y^3\exp\left(- \beta^{-1}L^{3/2}y^2\right)dy
=\int_{L^{-\alpha}}^{L^{-\alpha}}y\exp\left(- \beta^{-1}L^{3/2}y^2\right)dy=0.
\]
The two integrals resulting from the~$y^2$ and $y^4$-terms in~\eqref{E:refi} are easily calculated; they yield
a relative contribution of~$L^{-1/2}$, which merges with the term
$c_5 L^{-1/2}$. Hence we see that the absolute error term from the local expansion is 
\[
  x^{-A_{3}}L^{a/c^{2}-3/4}\exp(2\beta L^{1/2}) \times O(L^{-1/2}).
\]
%Now it is not hard to see that formula~(\ref{E:0}) follows from the previous estimates.
This completes the proof of Theorem~\ref{T:main}.
\begin{remark} \rm
Note that the preceding argument can be extended by taking more terms in the local
expansion of the integrand. A full asymptotic expansion in descending powers of $L=\log x$
can thus be obtained, which replaces the error term~$(1+O((\log x)^{-1/2}))$
in~\eqref{E:0} by
\[
  1 + C_1 (\log x)^{-1/2} + C_2 (\log x)^{-3/4} + \dots + O((\log x)^{-m/4})
\]
with some constants~$C_k$ and arbitrarily large~$m$.
This is a typical feature of the saddle point method (see~\cite{FlSe09}, Section~VIII.3).
\end{remark}
%
%\begin{remark}\label{R:const} \rm It follows from~(\ref{E:fin}) that for a correlated Heston model, the constants~$A_1$, $A_2$, and~$A_3$  
%are given by the following formulas:
%\begin{align*}
%A_{1}&=\frac{\exp \left( \Gamma \right) }{2\pi }\sqrt{\pi }\beta ^{1/2-2a/c^{2}}  \\
%&=\frac{1}{2\sqrt{\pi }}\exp \left( v_{0}\frac{b+s_+\rho c}{c^{2}}\right) \left(
%2v_{0}\right) ^{1/4-a/c^{2}}c^{-1/2+2a/c^{2}}\sigma ^{-a/c^{2}-1/4}, \\
%A_{2}&=2\beta =2\frac{\sqrt{2v_{0}}}{c\sqrt{\sigma }}, \\
%A_{3}&=s_++1.
%\end{align*}
%Observe that~$s^{\ast}$ and $\sigma =-\partial _{s}T^{\ast}|_{s^{\ast}}$, introduced in
%Definition~\ref{def:s_+}, depend on
%the correlation parameter~$\rho$. In Appendix II, we verify that the special case~$\rho=0$
%of the constants~$A_2$ and~$A_3$ defined in the present remark agrees
%with the corresponding constants found in~\cite{GuSt09}
%for uncorrelated Heston models.
%\end{remark}
%
\begin{remark} \rm
  By a standard result on integrating functions of regular variation~\cite[Proposition~1.5.10]{BiGoTe87},
  formula~\eqref{E:0} yields the estimate
  \[
    \mathbb{P}[S_T > x] = \frac{A_1}{A_3-1} x^{-A_3+1} e^{A_2\sqrt{\log x}}
      (\log x)^{-3/4+a/c^2} \bigl( 1+O( ( \log x) ^{-1/2}) \bigr),
  \]
  as $x\to\infty$,
  for the tail of the distribution of~$S_T$. Note that the main factor~$x^{-A_3+1}$ has been obtained
  by Dr{\u{a}}gulescu and Yakovenko~\cite[Section~6]{DrYa02}.
\end{remark}

\begin{remark}\label{rem:zero} \rm We briefly discuss the behavior of the Heston density~$D_T(x)$ near zero.
  Define the lower critical moment by
  \begin{equation*}
    s_{-}:=\inf \left\{ s\leq0 : E[S_{T}^{s}]<\infty \right\},
  \end{equation*}
  and the corresponding slope and curvature by
  \begin{equation*}
    \sigma_{-} := \partial _{s}T^{\ast }|_{s_{-}} \geq 0 \qquad \text{and} \qquad
    \kappa_{-}:=\partial _{s}^2T^{\ast }|_{s_{-}}.
  \end{equation*}
  As $x\downarrow 0$, the integrand in~\eqref{E:i1} has a saddle point that approaches
  the singularity $s_{-}+1$ at a speed of $(-\log x)^{-1/2}$. All steps of the subsequent
  analysis precisely parallel the case $x\to\infty$ treated above. The net result is
 \begin{equation}
    D_T(x) = B_1 x^{B_3} e^{B_2 \sqrt{-\log x}} (-\log x)^{a/c^2-3/4}
      \bigl(1+O((-\log x)^{-1/2})\bigr) 
 \label{E:est}
\end{equation}
as $x\downarrow 0$,  where
  \begin{align*}
    B_{3} &=-(s_{-}+1), \qquad B_{2} =2\frac{\sqrt{2v_{0}}}{c\sqrt{\sigma_{-} }},  \\
    B_1 &= \frac{1}{2\sqrt{\pi }} \left( 2v_{0}\right) ^{1/4-a/c^{2}}c^{2a/c^{2}-1/2}\sigma_{-}^{-a/c^{2}-1/4} \notag \\
      & \qquad \times  \exp\left( -v_0 \left( \frac{b+s_{-} \rho c}{c^2}+\frac{\kappa_{-}}{c^2\sigma_{-}^2} \right) 
      -\frac{aT}{c^2}(b+c\rho s_{-}) \right) \notag \\
      & \qquad \times
      \left(\frac{2\sqrt{b^2+2bc\rho s_{-}+c^2 s_{-}(1-(1-\rho^2)s_{-})}}{c^2 s_{-}(s_{-}-1)
      \sinh \frac12\sqrt{b^2+2bc\rho s_{-}+c^2 s_{-}(1-(1-\rho^2)s_{-})} } \right)^{2a/c^2}.
  \end{align*}%
\end{remark}

  \begin{figure}[!h]
   \centering
   \includegraphics[width=8truecm]{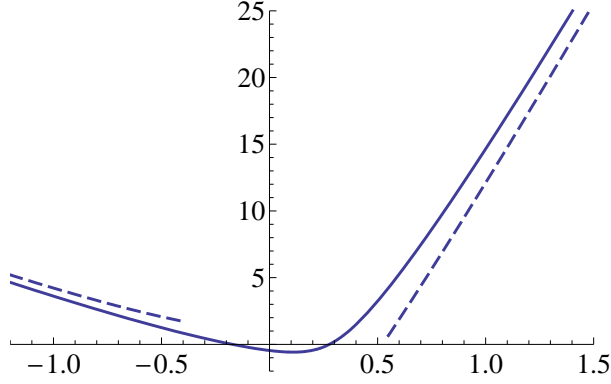}
   \caption{$-\log D^{\log}_T(x)$ with its asymptotic approximations,
    where~$D^{\log}_T$ is the density of $\log S_T$.}
    \label{fig:log-spot}
 \end{figure}

\begin{remark} \rm
  The density~$D^{\log}_T$ of the log-spot price~$\log S_T$ is given by
  \[
    D^{\log}_T(x) = e^x D_T(e^x).
  \]
  Its asymptotics readily follow from~\eqref{E:0} and~\eqref{E:est}:
  \[
    D^{\log}_T(x) = A_1 e^{-(A_3-1)x} e^{A_2\sqrt{x}} x^{a/c^2-3/4} \bigl(1+O(x^{-1/2})\bigr),
      \qquad x\to\infty,
  \]
  and
  \[
    D^{\log}_T(x) = B_1 e^{-(B_3+1)|x|} e^{B_2\sqrt{|x|}} |x|^{a/c^2-3/4} \bigl(1+O(|x|^{-1/2})\bigr),
      \qquad x\to\ -\infty.
  \]
  Figure~\ref{fig:log-spot} shows the numerical fit of these approximations, using a set
  \begin{align}
    & a =\bar{v}\lambda, \quad b = -\lambda, \quad c = 0.2928, \quad v_0= 0.0654,\quad  \rho = -0.7571, & \notag \\
    & \bar{v} = 0.0707,\quad  \lambda = 0.6067 & \label{eq:market}
  \end{align}
  of typical market parameters~\cite{ScSiTi04}.
\end{remark}

\section{Call pricing functions and smile asymptotics}
\label{se:smile}

Recall that our main result (Theorem~\ref{T:main}) is the following asymptotic formula for the stock price distribution density 
in a correlated Heston model with $S_0=1$:
\begin{equation}
  D_T(x)=A_1x^{-A_3}e^{A_2\sqrt{\log x}}(\log x)^{-\frac{3}{4}+\frac{a}{c^2}}
    \bigl(1+O((\log x)^{-\frac12})\bigr)
\label{E:a0}
\end{equation}
as $x\rightarrow\infty$. In the present section we will characterize the asymptotic behavior of the call pricing function $K\mapsto C(K)$ 
in such a model, and then prove Theorem~\ref{thm:impl vol}. The following formula is a generalization of a similar result obtained for uncorrelated Heston models in~\cite{Gu10}:
\begin{align}
C(K)&=\frac{A_1}{\left(-A_3+1\right)\left(-A_3+2\right)}K^{-A_3+2}e^{A_2\sqrt{\log K}}(\log K)^{-\frac{3}{4}+\frac{a}{c^2}} \nonumber \\
&\quad\times\left(1+O\left((\log K)^{-\frac{1}{4}}\right)\right)
\label{E:a1}
\end{align}
as $K\rightarrow\infty$. Formula~\eqref{E:a1}  
follows from~(\ref{E:a0}), Theorem 7.1 in~\cite{Gu10}, and Remark 6.1 in~\cite{Gu10}.
Note that $A_3> 2$. 

We will next use the tail-wing formulas obtained in~\cite{Gu10} to study the asymptotic behavior of the Black-Scholes implied volatility 
$K\mapsto\sigma_{BS}(K,T)$ in a correlated Heston model in the case where the maturity~$T$ is fixed and the strike~$K$ approaches infinity or zero.
The following statement was established in~\cite{Gu10}, Section~7.
Suppose that the stock price density~$D_T$ 
in a general stock price model satisfies the condition
\begin{equation}\label{E:estim}
  c_1 x^{-\xi}h(x)\le D_T(x)\le c_2 x^{-\xi}h(x)
\end{equation}
for all large~$x$, where $\xi> 2$, $h$ is a slowly varying function, and~$c_1$ and~$c_2$ are positive constants. 
%Suppose that the stock price density
%\begin{equation}\label{E:estim}
%  D_T(x) \sim x^{-\xi}h(x)
%\end{equation}
%in a general stock price model varies regularly at infinity, where
%$\xi> 2$, and $h$ is a slowly varying function. 
Then for every positive function~$\varphi$ on $(0,\infty)$
with $\lim_{x\rightarrow\infty}\varphi(x)=\infty$, we have the following:
\begin{align}
&\sigma_{BS}(K,T)\frac{\sqrt{T}}{\sqrt{2}}=\sqrt{\log K+\log\frac{1}{K^2D_T(K)}-\frac{1}{2}\log\log\frac{1}{K^2D_T(K)}} \nonumber \\
&\quad-\sqrt{\log\frac{1}{K^2D_T(K)}-\frac{1}{2}\log\log\frac{1}{K^2D_T(K)}}
+O\left(\left(\log K\right)^{-\frac{1}{2}}\varphi(K)\right) \nonumber \\
&=\sqrt{\log K+\log\frac{1}{K^{-\xi+2}h(K)}-\frac{1}{2}\log\log\frac{1}{K^{-\xi+2}h(K)}} \nonumber \\
&\quad-\sqrt{\log\frac{1}{K^{-\xi+2}h(K)}-\frac{1}{2}\log\log\frac{1}{K^{-\xi+2}h(K)}} \nonumber \\
&\quad+O\left(\left(\log K)\right)^{-\frac{1}{2}}\varphi(K)\right)
\label{E:eq}
\end{align}
as $K\rightarrow\infty$. 

A similar assertion holds for small values of the strike price (see~\cite{Gu10}, Section~7). It can be formulated as follows: 
Suppose that the stock price density~$D_T$ is such that  
\begin{equation}
c_1 x^{\gamma}h(x^{-1}) \leq D_T(x) \leq c_2 x^{\gamma}h(x^{-1})
\label{E:svs}
\end{equation}
for all sufficiently small $x>0$,
where $\gamma>-1$, $h$ is a slowly varying function, and~$c_1$ and~$c_2$ are positive constants. 
Let~$\tau$ be a positive function on $(0,\infty)$ with 
$\lim_{K\rightarrow 0}\tau(K)=\infty$. 
Then
\begin{align}
&\sigma_{BS}(K,T)\frac{\sqrt{T}}{\sqrt{2}}=\sqrt{\log\frac{1}{K^2D_T(K)}-\frac{1}{2}\log\log\frac{1}{KD_T(K)}} \nonumber \\
&\quad-\sqrt{\log\frac{1}{KD_T(K)}-\frac{1}{2}\log\log\frac{1}{KD_T(K)}}
+O\left(\left(\log\frac{1}{K}\right)^{-\frac{1}{2}}\tau(K)\right) \nonumber \\
&=\sqrt{\log\frac{1}{K^{\gamma+2}h(K^{-1})}
-\frac{1}{2}\log\log\frac{1}{K^{\gamma+1}h(K^{-1})}} \nonumber \\
&\quad-\sqrt{\log\frac{1}{K^{\gamma+1}h(K^{-1})}-\frac{1}{2}\log\log\frac{1}{K^{\gamma+1}h(K^{-1})}} \nonumber \\
&\quad+O\left(\left(\log\frac{1}{K}\right)^{-\frac{1}{2}}\tau(K)\right)
\label{E:eqo}
\end{align}
as $K\to 0$.
\begin{remark}\label{R:Lee} \rm
The asymptotic formulas in (\ref{E:eq}) and (\ref{E:eqo}) are equivalent to similar formulas with 
$\varphi(K)=1$ and $\tau(K)=1$, respectively. Indeed, if for some function $f$ and all functions $g$, which tend to infinity, 
we have $f(K)=O(g(K))$ 
as $K\rightarrow\infty$, then $f(K)=O(1)$ as $K\rightarrow\infty$. This can be shown as follows. If the function $f$ is not bounded near infinity,
then there exists a sequence $K_n\uparrow\infty$ such that $f(K_n)\ge 2^n$ for all $n\ge 1$. 
Put $g(K_n)=n$, and define the function $g$ by linear interpolation. 
Then $g(K)\rightarrow\infty$ as $K\rightarrow\infty$, but $f(K)\neq O(g(K))$ as $K\rightarrow\infty$. The proof for $K\rightarrow 0$ is similar.
The authors thank Roger Lee for bringing this simple fact to their attention.
\end{remark}

Now let us apply (\ref{E:eq}) and (\ref{E:eqo}) to the Heston model.
It is easy to see from~\eqref{E:a0} that~\eqref{E:estim} holds with  $\xi=A_3$ and the slowly
varying function
\[
  h(x)=e^{A_2\sqrt{\log x}}(\log x)^{a/c^2 - 3/4}.
\]
It follows from~\eqref{E:eq} and Remark~\ref{R:Lee} that
\begin{align}
&\sigma_{BS}(K,T)\frac{\sqrt{T}}{\sqrt{2}} \nonumber \\
&=\sqrt{\left(A_3-1\right)\log K-A_2\sqrt{\log K}-\left(\frac{a}{c^2}-
\frac{3}{4}\right)\log\log K-\frac{1}{2}\log\log \frac{1}{K^{-A_3+2}h(K)}}
\nonumber \\
&-\sqrt{\left(A_3-1\right)\log K-A_2\sqrt{\log K}-\left(\frac{a}{c^2}-
\frac{3}{4}\right)\log\log K-\frac{1}{2}\log\log \frac{1}{K^{-A_3+2}h(K)}}
\nonumber \\
&\quad+O\left((\log K)^{-\frac{1}{2}}\right)
\label{E:proof1}
\end{align}
as $K\rightarrow\infty$. Next, using the mean value theorem, we see that it is possible 
to replace the term $\frac{1}{2}\log\log\frac{1}{K^{-A_3+2}h(K)}$ under the square roots in formula 
(\ref{E:proof1}) by the term $\frac{1}{2}\log\log K$. Therefore,
\begin{align}
&\sigma_{BS}(K,T)\frac{\sqrt{T}}{\sqrt{2}} \nonumber \\
&=\sqrt{\left(A_3-1\right)\log K-A_2\sqrt{\log K}-\left(\frac{a}{c^2}-
\frac{1}{4}\right)\log\log K}
\nonumber \\
&\quad-\sqrt{\left(A_3-1\right)\log K-A_2\sqrt{\log K}-\left(\frac{a}{c^2}-
\frac{1}{4}\right)\log\log K}
\nonumber \\
&\quad+O\left((\log K)^{-\frac{1}{2}}\right)
\label{E:proof2}
\end{align}
as $K\rightarrow\infty$. Since $\sqrt{1-h}=1-\frac{1}{2}h+O\left(h^2\right)$ as $h\rightarrow 0$, formula
(\ref{E:proof2}) implies that
\begin{align}
&\sigma_{BS}(K,T)\frac{\sqrt{T}}{\sqrt{2}}=\left(\sqrt{A_3-1}-\sqrt{A_3-2}\right)\sqrt{\log K}\\
&\quad+\frac{A_2}{2}\left(\frac{1}{\sqrt{A_3-2}}-\frac{1}{\sqrt{A_3-1}}\right) \nonumber \\
&\quad+\frac{1}{2}\left(\frac{a}{c^2}-\frac{1}{4}\right)\left(\frac{1}{\sqrt{A_3-2}}
-\frac{1}{\sqrt{A_3-1}}\right)\frac{\log\log K}{\sqrt{\log K}}+O\left((\log K)^{-\frac{1}{2}}\right)
\label{E:proof3}
\end{align}
as $K\rightarrow\infty$. Next, using (\ref{E:proof3}),
we obtain the expansion~\eqref{ImplVarExp}
for the implied volatility $k\mapsto\sigma _{BS}(k,T)$, considered as a function of the
forward-log-in-moneyness $k=\log K$.
Theorem~\ref{thm:impl vol} is thus proved.
In the case where $\rho=0$, formula~\eqref{ImplVarExp} was obtained in~\cite{GuSt09} (see~\cite{GuSt09} and~\cite{Gu10} for more details).
Note that already the leading order term
\begin{equation*}
\sigma _{BS}( k,T) \sqrt{T} \sim \beta _{1}k^{1/2}, \qquad k\to\infty,
\end{equation*}
gives very good numerical approximation results. % (see~\cite{BeFr08})
This term was obtained in~\cite{BeFr08}. As a ``$\limsup$''-statement, based on
Lee's moment formula, it appears already in~\cite{AnPi07}.

Let us denote by~$W_{BS}$ the Black-Scholes implied total variance defined by 
\[
  W_{BS}(k,T)=\sigma_{BS}(k,T)^2T.
\] 
Then formula~(\ref{ImplVarExp}) implies the following expansion for~$W_{BS}$:
\[
W_{BS}(k,T)=\beta_1^2k+2\beta_1\beta_2k^{1/2}+2\beta_1\beta_3\log k+O(\varphi(k))\quad\mbox{as}\quad k\rightarrow\infty,
\]
where~$\beta_1$, $\beta_2$, $\beta_3$, and~$\varphi$ are the same as in~\eqref{ImplVarExp}.

Similar reasoning can be used in the case where $k\rightarrow-\infty$. Put $\gamma=B_3$ and 
\[
  h(x)=e^{B_2\sqrt{\log x}}(\log x)^{a/c^2 - 3/4},
\]
where $B_2$ and $B_3$ are defined in Remark \ref{rem:zero}. In addition, fix a positive function~$\varphi$ on $\left(0,\infty\right)$
with $\lim_{x\rightarrow\infty }\varphi(x) =\infty$.
Then~(\ref{E:est}) shows that all the conditions, under which formula~(\ref{E:eqo}) holds,
are satisfied. Next, using~(\ref{E:eqo}) and simplifying, 
we obtain the following asymptotic formula for the implied volatility in the Heston model:
\begin{equation}
\sigma _{BS}(k,T) \sqrt{T}=\rho_{1}(-k)^{1/2}+\rho_{2}+\rho
_{3}\frac{\log(-k)}{(-k)^{1/2}}+O\left( \frac{\varphi(-k) }{(-k)^{1/2}}
\right) 
\label{E:a2zero}
\end{equation}
as $k\rightarrow -\infty$. The constants in~(\ref{E:a2zero}) are given by
\begin{align*}
&\rho_{1}=\sqrt{2}\left( \sqrt{B_{3}+2}-\sqrt{B_{3}+1}\right), \\
&\rho_{2}=\frac{B_{2}}{\sqrt{2}}\left( \frac{1}{\sqrt{B_{3}+1}}-\frac{1}{
\sqrt{B_{3}+2}}\right), \\
&\rho_{3}=\frac{1}{\sqrt{2}}\left( \frac{1}{4}-\frac{a}{c^{2}}\right)
\left( \frac{1}{\sqrt{B_{3}+2}}-\frac{1}{\sqrt{B_{3}+1}}\right).
\end{align*}
For the total implied variance, we have
\[
W_{BS}(k,T)=\rho_1^2(-k)+2\rho_1\rho_2(-k)^{1/2}+2\rho_1\rho_3\log(-k)+O(\varphi(-k))
\]
as $k\rightarrow-\infty$.

\section*{Appendix I:\ Tail estimates}

It is known~\cite{RoFeUt09,Lu07} that all the singularities of the Mellin transform $E[e^{(u-1)X_t}]$ of
the stock price density~$D_T$ in the Heston model are
located on the real line. Therefore, the function 
$u\mapsto e^{\phi\left(u-1,T\right)+v_{0}\psi(u-1,T)}$ is analytic everywhere in the complex plane
except the points of singularity on the real line. The next statement justifies the application of the Mellin
inversion formula in~\eqref{eq:integral}, and will be useful in the tail estimate for the saddle
point method. By symmetry, it clearly suffices to consider the upper tail $(\Im(u)>0)$.
\begin{lemma}
\label{le:exptailWithODE}Let $T>0$ and 
$1\le s_1\le\Re (s)\le s_2$. Then the following estimate holds as $\Im (s)\rightarrow \infty$:
\begin{equation*}
\left\vert e^{\phi(s,T) +v_{0}\psi(s,T)}\right\vert =O(e^{-C\Im(s)}),
\end{equation*}%
where the constant $C> 0$ depends on~$T$, $s_1$, $s_2$, and~$v_0$.
\end{lemma}
\begin{proof}
Let $s=\xi+iy$ and suppose $y> 0$. We will first estimate the function~$\psi$. 
Recall that
\begin{equation*}
\dot{\psi}=\frac{1}{2}\left( s^{2}-s\right) +%
\frac{c^{2}}{2}\psi ^{2}+b\psi + s\psi \rho c \quad \text{ with } \quad \psi(\xi,0)=0.
\end{equation*}%
Set $\psi=f+ig$ and $\gamma =-\left( b+  \xi \rho c \right)$. Then
$\gamma\geq 0$, and we have
\begin{align*}
\dot{f} &= \frac{1}{2}\left( \xi^{2}-y^{2}-\xi\right) +\frac{c^{2}}{2}\left(
f^{2}-g^{2}\right) -\gamma f, \quad f(s,0) =0, \\
\dot{g} &= \frac{1}{2}\left( 2\xi y-y\right) +c^{2}fg-\gamma g, \quad
g(s,0) =0.
\end{align*}%
Our goal is to show that there exists a positive continuously differentiable function $t\mapsto C(t)$ on $[0,T]$ such that
\begin{equation}
f(s,t)\leq -C(t)y,
\label{E:diff0}
\end{equation}
where $s=\xi+iy$, $1\le s_1\le \xi\le s_2$, and~$y$ is large enough. We first observe that~$f$ satisfies 
the differential inequality%
\begin{align}
\dot{f}&\leq\frac{1}{2}\left( \xi^{2}-y^{2}-\xi\right) +\frac{c^{2}}{2}
f^{2}-\gamma f \\
&\leq-\frac{1}{3}y^{2}+\frac{c^{2}}{2}f^{2}-\gamma f
\label{E:compare}
\end{align}
for $y> y_0$, where~$y_0$ depends only on~$s_1$ and~$s_2$.
Set 
\[
V\left(y,r\right)=-\frac{1}{3}y^{2}+\frac{c^{2}}{2}r^{2}-\gamma r.
\] 
Then~(\ref{E:compare}) can be rewritten as follows:
\begin{equation}
\dot{f}(s,t)\le V(y,f(s,t))
\label{E:diff1}
\end{equation} 
where $s=\xi+iy$.

We will next find a function~$C(t)$, $t\in[0,T]$ with $C(0)=0$, strictly positive for $t>0$,
and such that the function $F(y,t):=-C(t)y$ satisfies the differential inequality
\begin{equation}
V\left(y,F\right)\leq\dot{F}.
\label{E:diff2}
\end{equation}

Let us first suppose that such a function~$C$ exists. Then it is clear that given $s=\xi+iy$, the initial data 
$F(y,0)=f(s, 0)=0$ match. Now
we can use the ODE comparison results and derive from~(\ref{E:diff1}) and~(\ref{E:diff2}) that~(\ref{E:diff0}) holds, which implies 
the following estimate:
\begin{equation}
\left|e^{v_0\psi(s,T)}\right|=e^{v_0f(s,T)}\le e^{-v_0C(T)\Im(s)}
\label{E:diff3}
\end{equation}
for all $s=\xi+iy$ with~$y$ large enough and $s_1\le \xi\le s_2$.

We now look for the function~$C$ satisfying the equation
\begin{equation*}
\dot{C}(t)=-\gamma C(t)+\theta,
\end{equation*}%
where~$\theta$ is a positive constant, and $C(0)=0$.
The solution of this equation is given by
\[
  C(t) =
  \begin{cases}
    \theta \gamma^{-1}(1-e^{-\gamma t})& \text{if}\ \gamma>0, \\
    \theta t & \text{if}\ \gamma=0.
  \end{cases}
\]
It follows that
for $t\in (0,T]$,
\begin{equation*}
0<C(t)\leq T\theta.
\end{equation*}%
Next, choosing $\theta >0$ for which $-\frac{1}{3}+\frac{c^{2}}{2}T^{2}\theta ^{2}=-%
\frac{1}{4}$, we obtain
\begin{align}
V\left(y, F(y,t)\right)&\leq-\frac{1}{3}y^{2}+\frac{c^{2}}{2}T^{2}\theta
^{2}y^{2}+\gamma C(t)y \nonumber \\
&=-\frac{1}{4}y^{2}+\left( \theta -\dot{C}(t)\right) y \nonumber \\
&\leq-\dot{C}(t)y=\dot{F}(y,t).
\label{E:diff4}
\end{align}
In~(\ref{E:diff4}), $y$ is large enough
and depends only on~$\theta$, and hence on the model parameter~$c$ and on~$T$.
This shows that the function~$F$ satisfies the differential inequality in~(\ref{E:diff2}), and it follows that
estimates~(\ref{E:diff0}) and~(\ref{E:diff3}) hold. 

Finally, we note that
\begin{equation*}
\Re (\phi (s,T) )=a\int_{0}^{T}f(s,t) \leq ay\left(
-\int_{0}^{T}C(t)dt\right)=-ay\tilde{C}(T).
\end{equation*}
Therefore, for~$\Im(s)$ large enough, 
\begin{equation*}
\left\vert e^{\phi(s,T) +v_{0}\psi(s,T)
}\right\vert \leq \exp \left\{-\left(a\tilde{C}(T)+v_{0}C(T)\right)\Im (s)\right\}.
\end{equation*}
The proof of Lemma~\ref{le:exptailWithODE} is thus completed.
\end{proof}

\begin{lemma}\label{L:tail1}
\label{le:outer tail} If $B>0$ is any constant, then the portion of
the integral~\eqref{eq:integral1} where $\Im(u)>B$ is $O(x^{-A_{3}}\exp (\beta L^{1/2}))$.
(Recall that $L=\log x$.)
\end{lemma}

\begin{proof}
If $\tilde{B}> B$ is a sufficiently large positive constant, then it easily follows from Lemma~\ref{le:exptailWithODE} that 
\begin{align*}
\left|\int_{\hat{u}+i\widetilde{B}}^{\hat{u}+i\infty}e^{-uL+\phi(u-1) +v_{0}\psi(u-1) }du\right|& \leq
Cx^{-A_{3}}\exp (\beta L^{1/2})\int_{\widetilde{B}}^{\infty}e^{-Cy}dy \\
& =O\left(x^{-A_{3}}\exp (\beta L^{1/2})\right).
\end{align*}%
(The integral is clearly~$O(1)$.) Moreover, since the Mellin transform of~$D_T$ does not have singularities outside the real line
(see~\cite{Lu07}), we have 
\[
\left|\int_{\hat{u}+iB}^{\hat{u}+i\widetilde{B}}e^{-uL+\phi +v_{0}\psi }du\right|=O(e^{-\hat{u}L})=
O\left(x^{-A_3}\exp(\beta L^{1/2})\right).
\]
This completes the proof of Lemma~\ref{L:tail1}.
\end{proof}

Lemma~\ref{L:tail1} shows that the part of the tail integral where $\Im (u)>B$ is asymptotically much smaller
than the central part. We will next estimate the whole
tail integral.

\begin{lemma}
\label{le:tail} The following estimate holds for the tail integral: 
\begin{equation*}
\left|\int_{\hat{u}+iL^{-\alpha }}^{\hat{u}+i\infty}e^{-uL+\phi +v_0\psi }du\right|=
x^{-A_3} \exp\left(2 \beta L^{1/2} - \tfrac12 \beta^{-1} L^{3/2-2\alpha} +
O(\log L) \right).
\end{equation*}
\end{lemma}

\begin{proof}
We will prove that there exists a constant $B>0$ such that the absolute value of the part of
the tail integral where $L^{-\alpha}<\Im(u)<B$ equals
\begin{equation}  \label{eq:le tail est}
x^{-A_3} \exp\left(2 \beta L^{1/2} - \tfrac12 \beta^{-1} L^{3/2-2\alpha} +
O(\log L) \right).
\end{equation}
It suffices to establish this statement, since Lemma~\ref{le:outer tail} shows
that the absolute value of the integral from $\hat{u}+iB$ to $\hat{u}+i\infty$ is
asymptotically smaller than the expression in~\eqref{eq:le tail est}.
(Indeed: Dividing~\eqref{eq:le tail est} by $x^{-A_3}\exp(\beta L^{1/2})$ yields
$\exp(\beta L^{1/2} + O(L^{3/2-2\alpha}))$, which tends to infinity. Note that $3/2-2\alpha < 1/2$ by~\eqref{eq:y range}.)

It follows from Lemma~\ref{le:dev} and Remark~\ref{R:vazh} that for some constant $\gamma> 0$,
\begin{equation*}
e^{\phi (u-1,T)+v_{0}\psi (u-1,T)}=O\left(\exp \left( \frac{\beta ^{2}}{A_{3}-u}
-\gamma\log (A_{3}-u)\right)\right)
\end{equation*}%
as~$u$ tends to $u^{\ast }=s_++1=A_{3}$ inside the analyticity strip.
More verbosely, there exists a constant $C> 0$ such that for a sufficiently small number $B> 0$ and 
for all~$u$ in the analyticity strip with $|\Im (u)|<B$ and $\Re (u)>u^{\ast }-B$, we have
\begin{equation*}
|e^{\phi (u-1)+v_{0}\psi (u-1)}|\leq C|A_{3}-u|^{-\gamma}
\exp \left( \Re\left( \frac{\beta ^{2}}{A_{3}-u}\right) \right). 
\end{equation*}%
Hence
\begin{align*}
&\left|\int_{\hat{u}+iL^{-\alpha }}^{\hat{u}+iB}e^{-uL+\phi +v_{0}\psi }du\right| \\
&\leq Cx^{-A_{3}}\exp (\beta L^{1/2})
\int_{L^{-\alpha }}^{B}|A_{3}-(\hat{u}+iy)|^{-\gamma}\exp \left( \Re \left( \frac{\beta ^{2}}{A_{3}-(\hat{u}+iy)}
\right) \right) dy \\
&\leq Cx^{-A_{3}}\exp (\beta L^{1/2})L^{\gamma/2}%\left. 
\exp \left(\frac{\beta ^{2}(A_{3}-\hat{u})}{(A_{3}-\hat{u})^2+L^{-2\alpha}} \right) \\
&=Cx^{-A_{3}}\exp \left( 2\beta L^{1/2}-
\beta^{-1}L^{3/2-2\alpha }+O(\log L)\right) .
\end{align*}

We have used that the factor $|A_{3}-(\hat{u}+iy)|^{-\gamma}$ grows only like a
power of~$L$, since
\begin{equation*}
\beta L^{-\frac{1}{2}}=A_{3}-\hat{u}\leq |A_{3}-(\hat{u}+iy)|.
\end{equation*}%
Furthermore, the quantity
\begin{equation}\label{eq:Re recipr}
\Re \left( \frac{\beta ^{2}}{A_{3}-(\hat{u}+iy)}\right) 
=\frac{\beta ^{2}(A_{3}-\hat{u})}{(A_{3}-\hat{u})^2+y^2}
\end{equation}%
decreases w.r.t.~$|y|$. Therefore, the integral~$\int_{L^{-\alpha}}^B$ of~\eqref{eq:Re recipr}
can by estimated by the value of its integrand at~$L^{-\alpha}$ times the length of the integration path. The latter is
absorbed into~$C$, and the former is given by
\begin{align*}
\frac{\beta ^{2}(A_{3}-\hat{u})}{(A_{3}-\hat{u})^2+L^{-2\alpha}}&=\beta L^{1/2}-\frac{\beta L^{1/2}}
{\beta^2L^{2\alpha-1}+1} \\
&=\beta L^{1/2}-\beta^{-1}L^{3/2-2\alpha}+O(L^{5/2-4\alpha}).
\end{align*}
(This can also by obtained by plugging $y=L^{-\alpha }$ into
the singular expansion~\eqref{eq:recipr expan} computed above.)
Finally, we write the factor $%
L^{\gamma/2}$ as $\exp (O(\log L))$.
\end{proof}

\section*{Appendix II:\ Comparison of constants}
Since~$s_+$ is the order of the critical moment, it is not hard to see that if $\rho=0$, then
the constant~$A_3$ defined by $A_3=s_++1$ is the same as the constant~$A_3$ in~\cite{GuSt09}. 

We will next show that for $\rho=0$, the constant~$A_2$ defined in~\eqref{eq:A_i} is the same as the
corresponding constant in~\cite{GuSt09}. It follows from~\eqref{eq:A_i} and from~\eqref{E:s} that the constant~$A_2$ 
used in the present paper for $\rho=0$ satisfies
\begin{equation}
A_2^2=\frac{8v_0}{c^2\sigma}
\label{E:s7}
\end{equation}
with
\[
\sigma=\frac{\left(2s_+-1\right)\left[Tc^2s_+\left(s_+-1\right)-2b\right]}
{2s_+\left(s_+-1\right)\left[c^2s_+\left(s_+-1\right)-b^2\right]}.
\]

We will next turn our attention to the constant~$A_2$ in~\cite{GuSt09}. Lemmas 6.6 and 7.3 established in~\cite{GuSt09}
provide an explicit expression for this constant. First note that the quantity $r=r_{\frac{1}{2}T|b|}$ in~\cite{GuSt09} and
the quantity~$s_+$ in the present paper are related by
\begin{equation}
r=\frac{T}{2}\left[c^2s_+(s_+-1)-b^2\right]^{\frac{1}{2}}.
\label{E:s2}
\end{equation}
This follows from the formula for~$A_3$ in~\eqref{eq:A_i} and from Lemmas~6.6 and~7.3 in~\cite{GuSt09}. 

It was shown in~\cite{GuSt09}, Lemmas~6.5, 6.6, and~7.3 that the following formula holds: 
\[
  A_2=\frac{B\sqrt{2}}{T^{\frac{1}{4}}(8C+T)^{\frac{1}{4}}}
\]
with
\begin{align*}
B&=\frac{\sqrt{2T}}{c}\left(\frac{Tv_0\sin r}{2c^2\frac{T^2}{8r}\left|\left(1+\frac{1}{2}T|b|\right)\cos r-r\sin r\right|}\right)
^{\frac{1}{2}}\left(b^2+\frac{4}{T^2}r^2\right)^{\frac{1}{2}} \\
&=\frac{2\sqrt{2}\sqrt{v_0}\sqrt{r\sin r}}{c^2\left|\left(1+\frac{1}{2}T|b|\right)\cos r-r\sin r\right|^{\frac{1}{2}}}
\left(b^2+\frac{4}{T^2}r^2\right)^{\frac{1}{2}}
\end{align*}
and
\[
  C=\frac{T}{2c^2}\left(b^2+\frac{4r^2}{T^2}\right).
\]
Hence,
\[
A_2=\frac{4\sqrt{v_0}\sqrt{r\sin r}}{c^2\sqrt{T}\sqrt{2s_+-1}\left|\left(1+\frac{1}{2}T|b|\right)\cos r-r\sin r\right|^{\frac{1}{2}}}
\left(b^2+\frac{4}{T^2}r^2\right)^{\frac{1}{2}}.
\]
Here we use the formulas for~$A_3$ in~\eqref{eq:A_i} and in Lemma~7.3 in~\cite{GuSt09}. 
Since $r\cos r+\frac{1}{2}T|b|\sin r=0$ and formula~(\ref{E:s2}) holds, we get the following relation between the constant~$A_2$ in~\cite{GuSt09} 
and~$s_+$:
\begin{align*}
A_2&=\frac{4\sqrt{v_0}r}{c^2\sqrt{T}\sqrt{2s_+-1}\left[\frac{1}{2}T|b|\left(1+\frac{1}{2}T|b|\right)+r^2\right]^{\frac{1}{2}}}
\left(b^2+\frac{4}{T^2}r^2\right)^{\frac{1}{2}} \\
&=\frac{4\sqrt{v_0}\sqrt{s_+\left(s_+-1\right)}\left[c^2s_+\left(s_+-1\right)-b^2\right]^{\frac{1}{2}}}
{c\sqrt{2s_+-1}\left[Tc^2s_+\left(s_+-1\right)-2b\right]^{\frac{1}{2}}}.
\end{align*}
Therefore,
\begin{equation}
A_2^2=\frac{16v_0s_+\left(s_+-1\right)\left[c^2s_+\left(s_+-1\right)-b^2\right]}
{c^2\left(2s_+-1\right)\left[Tc^2s_+\left(s_+-1\right)-2b\right]}.
\label{E:s5}
\end{equation}
Next, comparing~(\ref{E:s7}) and~(\ref{E:s5}), we see that the constant~$A_2$ used in the present paper coincides 
with the corresponding constant in~\cite{GuSt09}.

%\begin{figure}[h]
% \includegraphics[width=6truecm]{fit_A3}
% \includegraphics[width=6truecm]{fit_A2}
% \includegraphics[width=6truecm]{fit_A1}
% \caption{Numerical checks for the constants $A_3$, $A_2$, $A_1$.}
% \label{fig:A_i}
%\end{figure}

\section*{Appendix III: Numerical results}

\begin{figure}[!h]
\begin{center}
\includegraphics[width=0.8\linewidth]{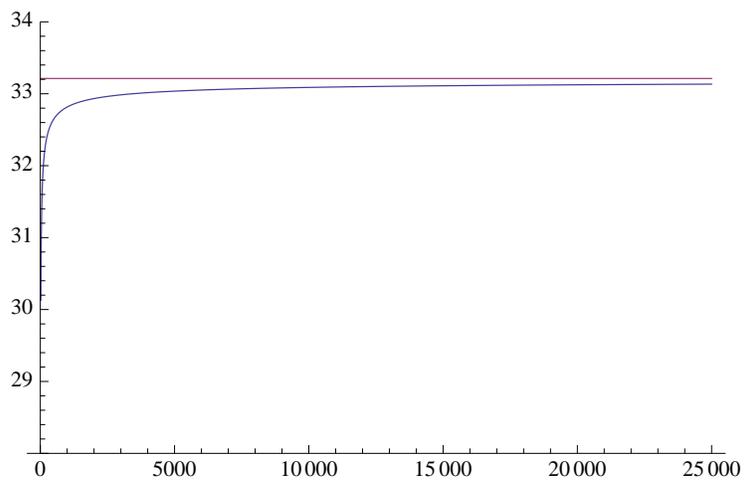}
\caption{Numerical check for the constant~$A_3$.}
\label{fig:A3}
\end{center}
\end{figure}

\begin{figure}[!h]
\begin{center}
\includegraphics[width=0.8\linewidth]{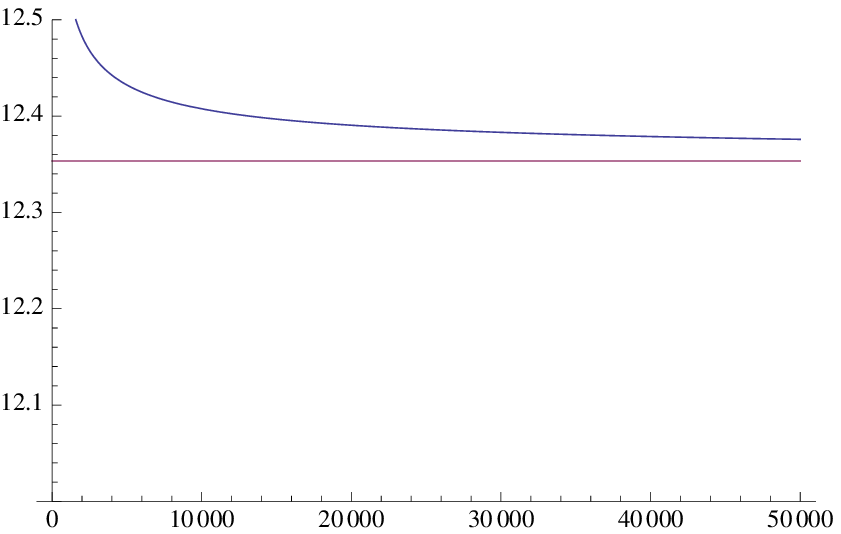}
\caption{Numerical check for the constant~$A_2$.}
\label{fig:A2}
\end{center}
\end{figure}

\begin{figure}[!h]
\begin{center}
\includegraphics[width=0.8\linewidth]{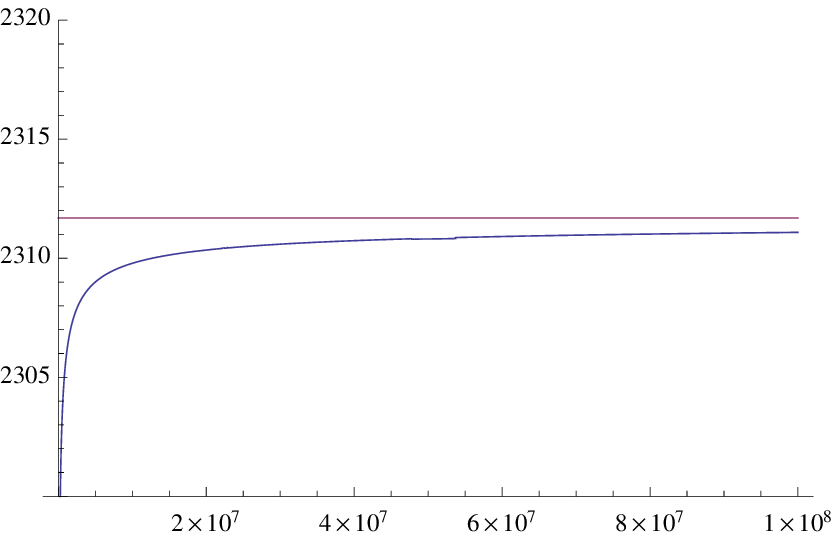}
\caption{Numerical check for the constant~$A_1$.}
\label{fig:A1}
\end{center}
\end{figure}

To conclude we illustrate the accuracy of~\eqref{E:0} by a numerical example, and show plots of
the corresponding smile approximations.
We will use the parameter values~\eqref{eq:market}.
%\begin{align*}
%& a =\bar{v}\lambda, \quad b = -\lambda, \quad c = 0.2928, \quad v_0= 0.0654,\quad  \rho = -0.7571, & \\
%& \bar{v} = 0.0707,\quad  \lambda = 0.6067. &
%\end{align*}
Note that~\eqref{E:0} implies that
\begin{align}
  &-\frac{\log D_T(x)}{\log x}  \to A_3 \approx 33.2124, \label{eq:A3 lim} \\
  &\frac{\log(x^{A_3} D_T(x))}{\sqrt{\log x}} \to A_2 \approx 12.3533, \label{eq:A2 lim} \\
  & \frac{x^{A_3} D_T(x)}{e^{A_2\sqrt{\log x}}(\log x)^{a/c^2-3/4}} \to A_1 \approx 2311.69, \label{eq:A1 lim}
\end{align}
as $x\to\infty$.
Figures~\ref{fig:A3}--\ref{fig:A1} plot the left- and right-hand sides of~\eqref{eq:A3 lim}--\eqref{eq:A1 lim},
with~$\log x$ on the horizontal axis. The density~$D_T$ was evaluated by numerical integration
of~\eqref{eq:integral}, using the explicit expressions~\cite{RoFeUt09, He93} for~$\phi$ and~$\psi$.

Finally, to show the accuracy of the smile asymptotics, we plot the smile together with the asymptotic approximations. This is done by simply matching Heston prices with Black-Scholes prices by means of a root-finding procedure.
To evaluate the Heston prices (with initial stock price $S_0 = 1$) we use Lee's formula~\cite{Le04b}
\[
C(T,k) = \frac{e^{-\alpha k}}{\pi}\int_0^\infty \Re{\Bigl(\frac{e^{-i uk}\phi (u-i(\alpha +1),T)}{\alpha^2+\alpha -u^2+i(2 \alpha+1)u}\Bigr)} \, du,
\]
where $k$ is again the log-strike and~$\alpha$ is a ``damping constant'' which we are free to choose, noting only that for $\alpha >0$ this formula gives us call prices whereas for $\alpha <-1$ we get the prices of the respective puts. To optimize our results, we will use (following Lee) call options for the out-of-the-money strikes, and  put options for the in-the-money strikes, both with maturity $T=1$. As a good choice for the damping constant~$\alpha$ we suggest $\alpha=29.1$ for the calls and $\alpha=-4.4$ for the puts.

\begin{figure}[!h]
\begin{center}
\includegraphics[width=0.8\linewidth]{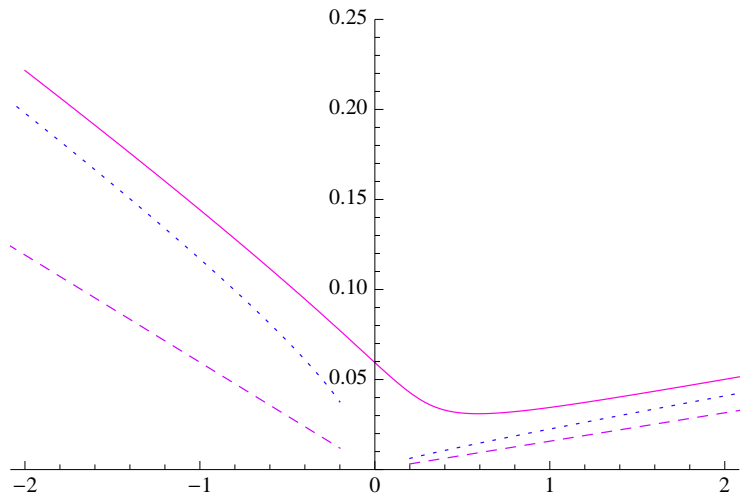}
\caption{Implied variance $\sigma(k,1)^2$ in terms of log-strikes compared to the first order (dashed) and third order (dotted) approximations.}
\label{fig:smile}
\end{center}
\end{figure}

The respective Black-Scholes prices are calculated by the Black-Scholes formula, evaluating the cumulative density function of the normal distribution by straightforward numerical integration.\footnote{We thank Roger Lee for helpful comments on this numerical
evaluation.}
To get good results for deep in-the-money/out-of-the-money options, we use as starting point for the root-finding procedure the value given by our third order approximation. In the numerical example this leads to a stable evaluation of the smile in a quite large interval, e.g.\ log-strikes ranging from~${-14}$ to~$24$. The results, compared with the first- and third order asymptotics, are found in Figure~\ref{fig:smile}. There, the log-strike
is confined to the (more realistic) interval $[-2,2]$.

\clearpage

\clearpage

%\nocite{Ga04,AvBoBuFr03,AvZh98,BeBuFl04,FeYa83,Ga06,HL05,HaKuLeWo02,HaLeWo05}
\nocite{Ga06}

\bibliographystyle{acm}
\bibliography{FrGeGuSt}

\end{document}